\documentstyle[prd,aps,preprint]{revtex}
\newcommand\ee{\end{equation}}
\newcommand\be{\begin{equation}}
\newcommand\eea{\end{eqnarray}}
\newcommand\bea{\begin{eqnarray}}

\def\so{S_1}
\def\st{S_3}
\def\stb{\overline{S}_3}
\def\se{S_8}
\def\see{S'_8}
\def\sf{S_{15}}
\def\sfp{S'_{15}}

\def\vo{|S_1|^2}
\def\vt{|S_3|^2}
\def\vtb{|\overline{S}_3|^2}
\def\ve{|S_8|^2}
\def\vee{|S'_8|^2}
\def\vf{|S_{15}|^2}
\def\vfp{|S'_{15}|^2}

\newcommand\Mpl{M_{\rm Pl}}

\newcommand\lsim{\mathrel{\rlap{\lower4pt\hbox{\hskip1pt$\sim$}}
    \raise1pt\hbox{$<$}}}
\newcommand\gsim{\mathrel{\rlap{\lower4pt\hbox{\hskip1pt$\sim$}}
    \raise1pt\hbox{$>$}}}

\def\dslash{\not{\hbox{\kern-2pt $\partial$}}}
\def\Dslash{\not{\hbox{\kern-4pt $D$}}}
\def\Oslash{\not{\hbox{\kern-4pt $O$}}}
\def\Qslash{\not{\hbox{\kern-4pt $Q$}}}
\def\pslash{\not{\hbox{\kern-2.3pt $p$}}}
\def\kslash{\not{\hbox{\kern-2.3pt $k$}}}
\def\qslash{\not{\hbox{\kern-2.3pt $q$}}}

\def\eeq{\end{equation}}
\def\beq{\begin{equation}}

\begin{document}
\topmargin-2.5cm
%
\begin{titlepage}
\begin{flushright}
CERN-TH/97-7\\
OUTP-99-01-P\\
IEM-FT-172/98
\end{flushright}
\vskip 0.3in
\begin{center}
{\Large\bf $D$-term Inflation in Superstring Theories}

\vskip .5in
{\bf J.R. Espinosa~$^{\dagger,}$\footnote{ E-mail: 
{\tt
espinosa@mail.cern.ch.}}},
{\bf A. Riotto~$^{\dagger,}$\footnote{
On leave of absence from the Department of Physics, Theoretical 
physics,
University of Oxford, U.K.  Email:
{\tt riotto@nxth04.cern.ch.}}} and
{\bf G.G. Ross~$^{\S,}$\footnote{ E-mail: {\tt
g.ross1@physics.oxford.ac.uk.}}}

\vskip.35in

$^{\dagger}$~Theory Division, CERN, CH-1211 Geneva 23, Switzerland.\\

\vskip 0.3cm

$^{\S}$~Department of Physics, Theoretical physics, University of 
Oxford,
U.K.

\vskip 0.3 cm

\end{center}
\vskip1.3cm
\begin{center}
{\bf Abstract}
\end{center}
\begin{quote}

An inflationary stage dominated by a $D$-term avoids the slow-roll
problem of inflation in supergravity  and may emerge in theories with a
non-anomalous or anomalous $U(1)$ gauge symmetry. The most intriguing 
and commonly invoked possibility is that the Fayet-Iliopoulos $D$-term  triggering inflation  is the one emerging in superstring theories. We discuss the complications  one has to face when trying to build up a successful
$D$-term inflationary scenario in  superstring models. In particular, we
show that the ``vacuum shifting''  phenomenon of string theories is usually very efficient even in the early Universe, thus preventing inflation from
taking place. On the other hand, when $D$-term inflation is free to 
occur, the presence of a plethora of fields and several non-anomalous 
additional abelian symmetries in string theories may help in  reconciling   the
 value of the Fayet-Iliopoulos  $D$-term required by the COBE 
normalization with  the value predicted by string theories. We also show that in superstring $D$-term inflation gravitinos are likely to pose no 
cosmological problem.

\end{quote}
\vskip1.cm
\end{titlepage}
\setcounter{footnote}{0}

\newpage
%

\noindent

\newpage
\setcounter{page}{1}
\baselineskip=20pt

{\bf 1.}~~The flatness and the horizon problems of the standard big 
bang
cosmology are elegantly solved if during the evolution of the early
Universe the energy density happens to be dominated by some form of
vacuum energy and comoving scales grow quasi-exponentially \cite{guth}.
An inflationary stage is also required to dilute any undesirable
topological defects left as remnants after some phase transition taking
place at early epochs. The vacuum energy driving inflation is generally
assumed to be associated with some scalar field $\phi$, the inflaton,
which is initially displaced from the minimum of its potential. As a
by-product, quantum fluctuations of the inflaton field may be the seeds
for the generation of structure. The level of density and
temperature fluctuations observed in the present Universe,
$\delta\rho/\rho\sim 10^{-5}$, require the inflaton potential to be
extremely flat. This means that the couplings of the inflaton field to
the other degrees of freedom cannot be too large: large couplings 
induce
large loop corrections to the inflaton potential, spoiling its 
flatness.
This is the main reason why inflation is more natural in the context of
supersymmetric theories. Introducing very small parameters to ensure 
the
extreme flatness of the inflaton potential seems very fine-tuned in 
most
non-supersymmetric theories, while this naturalness is achieved in
supersymmetric models. The nonrenormalization theorems in exact global
supersymmetry guarantee that we can fine-tune any parameter at the
tree-level and this fine-tuning will not be destabilized by radiative
corrections at any order in perturbation theory \cite{grisaru79}.

There is, however, a severe problem one has to face when dealing with
supersymmetric inflation model building in the context of 
supersymmetric
theories. The generalization of supersymmetry from a global to a local
symmetry automatically incorporates gravity and, therefore, inflation
model building must be considered in the framework of supergravity
theories. The supergravity potential $V$ consists of two pieces, the
so-called
  $D$-term and $F$-term.  For models where the $D$-term vanishes, the
slow-roll parameter $\eta=\Mpl^2 V''/V$, where $\Mpl\simeq 2.4\times
10^{18}$ GeV is the reduced Planck scale, generically receives various
contributions of order $\pm 1$ \cite{dinefisch}. This is the so-called $\eta$-problem of
supergravity theories:  there are contributions of order $\pm H^2$ to
the mass-squared of every scalar field and the troublesome 
contributions
to $\eta$ may be regarded as contributions to the coefficient $m^2$ in
the expansion of the inflaton potential. Therefore, it is very
difficult naturally to implement a slow-roll inflation in the context 
of
supergravity. The problem basically arises since inflation, by
definition, breaks global supersymmetry because of a nonvanishing
cosmological constant (the false vacuum energy density of the
inflaton). In supergravity theories, supersymmetry breaking is
transmitted by gravity interactions and the squared mass of the 
inflaton
becomes naturally of order of $V/\Mpl^2\sim H^2$. The perturbative
renormalization of the K\"ahler potential is therefore crucial for the
inflationary dynamics due to a non-zero energy density which breaks
supersymmetry spontaneously during inflation. How severe the problem is
depends on the magnitude of $\eta$. If $\eta$ is not too small then its
smallness could be due to accidental cancellations. Having $\eta$ not
too small requires that the spectral index $n=1-6\epsilon+2\eta$
[$\epsilon=\frac{1}{2}\Mpl^2(V'/V)^2$ is another slow-roll parameter] 
be
not too small, so the observational bound $|n-1|<0.3$ is already
beginning to make an accident look unlikely.

{\bf 2.}~~Solutions to the $\eta$-problem already exist in the
literature \cite{lr}. Among them, $D$-term inflation seems to be 
particularly promising \cite{ewansgrav,bindvali,halyo}\footnote{It is
interesting to notice that $D$-term inflation may provide a solution to
the moduli problem too \cite{moduli}.}.  It is based on the observation
that, when the vacuum energy density is dominated by nonzero $D$-terms
and supersymmetry breaking is of the $D$-type, scalars get 
supersymmetry
soft breaking masses which depend only on their gauge charges. Scalars
charged under
 the corresponding gauge symmetry obtain a mass much larger than $H$,
while gauge singlet fields can only get masses from loop gauge
interactions.  In particular, if the inflaton field is identified with 
a
gauge singlet, its potential may be flat up to loop corrections and
supergravity corrections to $\eta$ from the $F$-terms are not present
since the latter vanish during inflation\footnote{See, however, Ref.
\cite{lyth} for further comments.}. The toy model adopted in Refs.
\cite{bindvali,halyo} contains three chiral superfields $S$, $\Phi_+$
and $\Phi_-$ with charges equal to $0$, $+ 1$ and $- 1$ respectively
under a $U(1)$ gauge symmetry.  The superpotential has the form 
\beq W =
\lambda S\Phi_+\Phi_-.  \end{equation} The scalar potential in the
global supersymmetry limit reads \begin{equation} V = \lambda^2 |S|^2
\left(|\phi_-|^2 + |\phi_+|^2 \right) + \lambda^2|\phi_+\phi_-|^2 + 
{g^2
\over 2} \left(|\phi_+|^2 - |\phi_-|^2 + \xi \right)^2 \end{equation}
where $\phi_{\pm}$ are the scalar fields of the supermultiplets
$\Phi_{\pm}$, $g$ is the gauge coupling and what is crucial is the
presence of the Fayet-Iliopoulos (FI) $D$-term $\xi>0$. The global
minimum is supersymmetry conserving, but the gauge group $U(1)$ is
spontaneously broken \begin{equation} \langle S \rangle = \langle 
\phi_+
\rangle = 0, ~~~ \langle \phi_-\rangle = \sqrt{\xi}.  
\end{equation}
However, if we minimize the potential, for fixed values of $S$, with
respect to other fields, we find that for $S > S_c = g 
\sqrt{\xi}/\lambda$,
the minimum is at $\phi_+ =\phi_- = 0$. Thus, for $S > S_c$
and $\phi_+ =\phi_- = 0$ the tree level potential has a vanishing
curvature in the $S$ direction and large positive curvature in the
remaining two directions $ m_{\pm}^2 = \lambda^2|S|^2 \pm g^2\xi$. For
arbitrarily large $S$ the tree level value of the potential remains
constant, $V = g^2\xi^2/2$, and $S$ plays the role of the
inflaton. As stated above, the charged fields get very large masses due
to the $D$-term supersymmetry breaking, whereas the gauge singlet field
is massless at the tree-level.

Along the inflationary trajectory $\phi_{\pm}=0$, $S\gg S_c$, all the
$F$-terms vanish and large supergravity corrections to the
$\eta$-parameter do not appear. Therefore , we do not need to make any
assumption about the structure of the K\"ahler potential for the
$S$-field: minimal $S^*S$ and non-minimal quartic terms in the K\"ahler
potential $(S^*S)^2$ (or even higher orders) do not contribute in the
curvature, since $F_S$ is vanishing during inflation.

Since the energy density is dominated by the $D$-term, supersymmetry is
broken and this amounts to splitting the masses of the scalar and
fermionic components of $\Phi_{\pm}$. Such splitting results in the
one-loop effective potential for the inflaton field
\begin{equation}
V_{{\rm 1-loop}} = {g^2 \over 2}\xi^2 \left( 1 + {g^2 \over 8\pi^2}
{\rm ln} {\lambda^2 |S|^2 \over \mu^2}\right), 
\end{equation}
where $\mu^2$ is the renormalisation scale. Equivalently,
\be
V_{{\rm 1-loop}} = {g^2(\mu^2=\lambda^2 |S|^2) \over 2}\xi^2,
\ee
where the loop-correction is absorbed in the gauge
coupling running with the supersymmetric one-loop RG
\be
\mu { dg^2\over d\mu}={1\over 16\pi^2}g^4\sum_i Q_i^2,
\ee
with the sum extending to all the fields in the model. Note that scale 
invariance
of the effective potential $dV/d\mu=0$ can be used to evaluate the
coefficient of the one-loop logarithmic contribution directly.

The end of
inflation is determined either by the failure of the slow-roll
conditions or when $S$ approaches $S_c$. COBE \cite{COBE} imposes the 
following
normalization
\begin{equation}
\label{c}
5.3\times 10^{-4} =
\left(\frac{V^{3/2}}{V'\Mpl^3}\right)_*
 \end{equation}
where the * denotes this is to be evaluated at the scale at which the 
relevant fluctuations leave the horizon. This can be written in the
equivalent form \begin{equation}
 \left(\frac{V^{1/4}}{\epsilon^{1/4}}\right)_*=8\times 10^{16}\:{\rm GeV}. 
\end{equation} More or less independently of the value of $|S|$ at the
end of inflation, this gives with the above potential
\beq \label{cobe}
\sqrt{\xi}_{{\rm COBE}}=6.6\times 10^{15}\: {\rm GeV}. 
\end{equation}
This normalization is independent of the gauge coupling
 constant $g$ in the toy model, but if we write the potential during 
inflation as
\begin{equation}
\label{cc}
V=V_0(1+c \:{\rm ln}|S|),
\end{equation}
it
depends on the numerical coefficient $c$
in  the one-loop potential and it scales like $c^{1/4}$. This, in turn, 
depends
upon the particle
content of the specific model under consideration through the RG beta 
function of
the gauge coupling, as explained above.

Notice that, if the theory contains an abelian $U(1)$ gauge symmetry
(anomalous or not), the FI $D$-term  \begin{equation} \xi \int d^4
\theta ~V = \xi D \end{equation} is gauge invariant and therefore
allowed by the symmetries and it may lead to $D$-type supersymmetry
breaking. It is important to notice that this term may be present in 
the
underlying theory from the very beginning. Successful $D$-term 
inflation models
 based on this observation have been constructed in the framework of 
Granf Unified
Theories (GUT's) \cite{dr}.

In the rest of the paper, however, we would like to focus on a more
 intriguing possibility that is usually invoked to motivate $D$-term 
inflation:
in string theories, there can be a $U(1)_A$ gauge
symmetries which is anomalous. This means that ${\rm Tr} {\bf Q}_A\neq
0$ where the trace is evaluated over the massless string
states. Indeed, string theory provides a different mechanism, the
Green-Schwarz mechanism \cite{u(1)A}, by which the anomaly may be
cancelled even though the trace is nonvanishing. Such a nonvanishing
trace leads to the appearance of a one-loop FI $D$-term of the form
\cite{fi} \begin{equation} \label{string} \xi =
\frac{g^2}{192\pi^2}\:{\rm Tr} {\bf Q}_A\:\Mpl^2. \end{equation} Then
$\sqrt{\xi}$ is expected to be of the order of the string scale,
$(10^{17}-10^{18})$ GeV or so.

{\bf 3.}~~There are still  many open questions related to $D$-term
inflation in the framework of superstring theories :

{\it --The Fayet-Iliopoulos $D$-term from string theories faces the 
COBE
normalization--}

Comparing the value of the FI $D$-term normalized according to COBE to
the value predicted by string theories, it is clear that the $\xi_{{\rm
COBE}}$ looks too small to be consistent with the value arising in many
compactifications of the heterotic string. Notice that this problem may
be exacerbated if cosmic strings are formed at the end of inflation
\cite{lr1}. Some level of flexibility may be allowed in the case in
which, in the strong coupling limit, the ten-dimensional $E_8\otimes
E_8$ heterotic string can be described as the compactification of an
eleven-dimensional theory known as M-theory \cite{mtheory}.
When the ten-dimensional heterotic coupling is large, the fundamental
eleven-dimensional mass parameter $M_{11}$ becomes of the order of the
unification scale and it might be that the value of the FI $D$-term may
be reduced to a value closed to the one required by COBE. However, to 
our
knowledge no explicit example has been constructed so far. It is also
clear that the whole issue is strictly related to another unsolved
problem in heterotic string theories, the stabilisation of the dilaton
field.  Since the dilaton potential most likely is strongly influenced
by the inflationary dynamics, the actual value of $\xi$ at the moment
 when observationally interesting scales crossed the horizon during
inflation might be quite different from the one "observed" today
\cite{sarkar,dvaliriotto,matsuda}.

{\it --Constructing a viable model--}

The presence of the FI $D$-term (\ref{string}) leads to the breaking of
supersymmetry at the one-loop order at very high scale, an option which
is not phenomenologically viable. The standard solution to this puzzle
is to give a nonvanishing vacuum expectation values (VEV's) to some of
the scalar fields which are present in the string model and are charged
under the anomalous $U(1)_A$. In such a way, the FI $D$-term is
cancelled and supersymmetry is preserved. In the context of string
theory, this procedure is called ``vacuum shifting'' since it amounts
to moving to a point where the string ground state is stable. While
maintaining the $D$- and $F$-flatness of the effective field theory,
such vacuum shifting may have important consequences for the
phenomenology of the string theory. Indeed, the vacuum shifting not
only breaks the $U(1)_A$, but may also break some other gauge 
symmetries
under which the fields which acquire a VEV are charged. This is because
the anomalous $U(1)_A$ is usually accompanied by a plethora of
nonanomalous abelian symmetries.

 The vacuum shifting can generate effective superpotential mass terms
for vector-like states that would otherwise remain massless or may even
be responsible for the soft mass terms of squarks and sleptons at the
TeV scale \cite{dudas}.

What is relevant for our considerations is that in string theories the
protection of supersymmetry against the effects of the anomalous
$U(1)_A$ is extremely efficient.  If we now apply a sort of ``minimal
principle'' \cite{dine,dvaliriotto} requiring that a successful 
scenario
of $D$-term inflation should arise from a ``realistic'' string model
leading to the $SU(3)_C\otimes SU(2)_L\otimes U(1)_Y$ gauge structure 
at
low energies, the requirement of cancellation of the FI $D$-term by the 
vacuum shifting
mechanism may (and usually does) represent a serious problem. In other
words, how can we guarantee that during inflation the FI $D$-term is
not cancelled by one of the many scalar fields which are charged under
the anomalous $U(1)_A$ and are not coupled to the inflaton? Does a
successful $D$-term inflationary scenario in string theory require many
inflatons to render the vacuum shifting mechanism inoperative? Is it 
possible
that the presence of several fields and non-anomalous $U(1)$'s may 
solve
the problem of the mismatch between the
 value of the Fayet-Iliopoulos  $D$-term required by the COBE 
normalization and 
 the value predicted by string theories? Do gravitinos or  other 
dangerous relics
pose any cosmological problem?
It is
clear that only a systematic analysis of specific models can answer 
these and similar questions. This requires the
identification of possible inflatons and $D$- and $F$-flat
directions for a large class of perturbative string vacua.  This
classification \cite{jose1} is a prerequisite to address systematically the issue of
$D$-term inflation in string theories as well as the phenomenological
issues at low energy \cite{jose1,jose2,florida}.

{\bf 4.}~~ As an illustrative example of the possible complications one
has to face in building up a successful
 model of $D$-term inflation  in the framework of  4D string models,
we consider the massless spectrum of a compactification on a Calabi-Yau
manifold with Hodge numbers $h_{1,1}$, $h_{2,1}$,  etc. The 
four-dimensional
gauge group is $SO(26)\times U(1).$ There are then $h_{1,1}$ 
left-handed
chiral supermultiplets transforming as $({\bf 26},\surd
\frac{1}{3}$)$\oplus ({\bf 1 },-2\surd \frac{1}{3})$ and $h_{2,1}$
supermultiplets transforming as $({\bf 26},-\surd \frac{1}{3}$)$\oplus
({\bf 1},2\surd \frac{1}{3}).$ In this case
the $U(1)$ is clearly anomalous. The one-loop D-term is given by

\begin{equation}
\xi =\frac{g^2\Mpl^2}{192\pi ^{2}}\sum_{i}n_{i}Q_{i}h_{i}=\frac{g^2 \Mpl^2}{192\pi
^{2}}\cdot 2 \cdot
\frac{24}{\surd 3}(h_{1,1}-h_{2,1})
 \label{as}
\end{equation}
where the sum is over the (positive helicity) states with $U(1)$ charge 
$
q_{i}$ and multiplicity $n_{i}.$ It is positive if the Euler
characteristic $
\chi =2(h_{1,1}-h_{2,1})>0,$ the sign being fixed by the dominant
contribution of the $SO(26)$ non-singlet fields. In addition we suppose 
the
model has a gauge singlet field $S$ which will play the role of the
inflaton. Further we assume that there is a discrete $R$-symmetry that 
ensures
$S$-flatness. These assumptions are quite ad hoc and in a realistic 
model we
would have to demonstrate the existence of such a field,  but we use this
simple example to illustrate another problem that must be overcome if 
one is
to obtain a realistic string model of $D$-term inflation.

With this field we may try to construct an inflationary potential along 
the
lines of Eq (2). In particular one may generate masses for the 
$h_{2,1}$
vectorlike combinations of the $SO(26)$ singlet and non-singlet fields 
via
the couplings in the superpotential of the form
\begin{equation}
W=\lambda S\left[({\bf 26},\surd \frac{1}{3})\cdot({\bf 26},-\surd
\frac{1}{3})+({\bf
1},-2\surd \frac{1}{3})\cdot({\bf 1},2\surd \frac{1}{3})\right].  
\label{mass}
\end{equation}
This leaves light $(h_{1,1}-h_{2,1})$ fields transforming as $({\bf 26}
,\surd \frac{1}{3}$)$\oplus ({\bf 1},-2\surd \frac{1}{3})$ and leaves
unchanged Eq. \ref{as}.  Only the $SO(26)$ singlet fields, $\phi
_{i},$ are now available to cancel the anomalous $D$-term and indeed 
their
tree-level couplings to $D$ are negative :$\sum_{i}$ $Q_{i}|\phi 
_{i}|^{2}<0$, 
as is expected if supersymmetry is not to be broken by the FI $D$-term. However
this prevents us from implementing $D$-term inflation because the 
scalar
potential dependence on the $\phi _{i}$ fields arises only through the
anomalous $D$-term of the form
\begin{equation}
\frac{g^{2}}{2}\left(\xi + \sum_{i}Q_{i}|\phi _{i}|^{2}\right)^{2}.
\end{equation}
The vacuum expectation values  of the fields $\phi _{i}$ will rapidly
flow to cancel the $D$-term preventing inflation from occurring.

This example illustrates the problem in implementing $D$-term inflation 
in a
string theory. It arises because the minimum of the potential should 
not
break supersymmetry through the anomalous $D$-term and so there must be
light fields (here the $\phi _{i})$ with the appropriate $U(1)$ charge 
to
cancel it. To implement $D$-term inflation these fields must acquire a 
mass
for large values of $S$ but this was not possible in this example 
because
the $\phi _{i}$ were protected by chirality from acquiring mass by 
coupling to
the $S$ field.

Thus we conclude that it is crucial to consider {\it all} fields with
non-trivial $U(1)$ quantum numbers when discussing the possible
inflationary potential in the framework of superstring theories.

{\bf 5.} 
We will consider now further examples to capture  other possible
aspects of  $D$-term inflation in superstring theories. For 
illustrative purposes, 
we will use the specific string models, discussed in \cite{CHL,AF1} 
whose space of
flat directions was recently analyzed in \cite{jose1}. The 
emphasis will
be on exploring the different possibilities that may be realized rather 
than
proposing  a working model of inflation.  In so doing we will often
restrict the analysis to some subset of the fields present in the model
and ignore the rest. In view of what we concluded above, this is not
consistent, but the examples that follow should only be considered
as toy models attempting to capture some of the stringy characteristics
one should expect  when trying to construct  a fully realistic model of 
$D$-term inflation
in superstring inspired scenarios. 

The presence of several (non-anomalous) additional $U(1)$ factors is a
generic property of string models.  For the discussion of $D$-term
inflation, the relevant objects are thus no longer single elementary 
fields but
rather multiple-field directions in field space along which the $D$-term 
potential of
the non-anomalous $U(1)$'s vanishes \cite{jose2}. These directions would be truly 
flat if
an anomalous $U(1)_A$ (or some $F$-terms) were not present.  To study
whether a given direction remains flat in the
presence of the anomalous $U(1)_A$, the important quantity is the 
anomalous
charge along the direction. If the sign of this charge is opposite to
that of the Fayet-Iliopoulos term, VEVs along the flat direction will
adjust themselves to cancel the FI term and give a zero potential.  If 
the
charge has the same sign of the FI term, the potential along that
direction rises steeply with increasing values of the field. The
interesting case
corresponds to zero anomalous charge, in which case the potential
along the given direction is flat and equals, at tree level,
$g_A^2\xi^2/2$. The parallelism with the toy model of section~{\bf 2} 
is
evident.

The condition $Q_A=0$ ensuring tree-level flatness is not by itself
sufficient. We must also require that the direction is stable for large
values of the field, that is, all masses deep in the inflaton direction
must be  positive (or zero). However the presence of the FI term in the
scalar potential can induce negative masses for those fields which have
a negative anomalous charge (recall we are taking $\xi>0$):
\be
\label{massxi}
\delta m_i^2 = g_A^2Q^A_i\xi.
\ee
To ensure that masses are positive in the end one can use $F$-term
contributions (to balance the negative FI-induced masses) coming from
superpotential terms of the generic form
\be
\label{wf}
\delta W=\lambda I'\Phi_+\Phi_-,
\ee
where $I'$ stands for some product of fields that enter the inflaton
direction while $\Phi_\pm$ do not. Fields of type $\Phi_+$ and
$\Phi_-$ which couple to the inflaton
direction in the  superpotential terms get a large $F$-term
mass, $ \lambda \langle I' \rangle$.

Consider the simplest example, a toy model with two chiral fields $S_1$ 
and
$S_2$ of opposite $U(1)$ charges, so that the direction 
$|S|=|S_1|=|S_2|$ can
play the role of the inflaton. Assume that deep in this direction 
($S\gg
\sqrt{\xi}$) the masses of all fields are positive (or zero) and thus 
no other VEVs are
triggered. Then we can minimize the $D$-term scalar 
potential\footnote{In writing
this potential we are assuming for simplicity that kinetic mixing of 
different
$U(1)$'s is absent. For this to be a consistent assumption the 
vanishing of
${\rm Tr}(Q_A Q_\alpha)$ and ${\rm Tr}(Q_\alpha Q_\beta)$ is a necessary
condition.}
\bea
V_D&=\frac{1}{2}g_A^2\left[ Q_1^A\left(|S_1|^2-|S_2|^2\right)+
\sum_iQ_i^A|\phi_i|^2+\xi\right]^2\nonumber\\
&+\frac{1}{2}\sum_\alpha g_\alpha^2\left[ 
Q_1^\alpha\left(|S_1|^2-|S_2|^2\right)+
\sum_iQ_i^\alpha|\phi_i|^2\right]^2,
\eea
[where $\alpha=1,...,n$ counts the additional D-term contributions of 
the non-anomalous $U(1)$'s] for
$S_1$ and $S_2$ only.

If $\xi=0$, $|S_1|=|S_2|$ is flat and necessarily stable, as 
$V=0$. For $\xi>0$
however, the flat direction is slightly displaced and lies at
\be
\delta S^2\equiv |S_1|^2-|S_2|^2=-\frac{g_A^2}{G_{11}^2}Q_1^A\xi,
\label{s2}
\ee
where $G_{ij}^2=g_A^2Q_i^AQ_j^A+\sum_\alpha
g_\alpha^2 Q_i^\alpha Q_j^\alpha$. This displacement is the result of 
the
destabilization effect of $\xi$ referred to above and occurs  when the
fields in
the inflaton direction carry anomalous charge: as the inflaton 
direction must have
zero
anomalous charge, the fields forming it have anomalous charges of 
opposite signs
and one of them will get a negative mass of the form (\ref{massxi}).
Notice that,  a term like (\ref{wf}) but with say $\Phi_-$ belonging
 to the inflaton direction cannot be used to stabilize inflaton fields,
because it
would spoil the $F$-flatness of the inflaton direction.

Taking into account this displacement, the value of the potential along 
the
inflaton direction is, at tree level
\be
V_0=\frac{1}{2}\frac{g_A^2}{G_{11}^2}\xi^2\sum_\alpha 
g_\alpha^2(Q_1^\alpha)^2
\equiv \frac{1}{2}g_A^2\xi_{eff}^2\leq\frac{1}{2}g_A^2\xi^2.
\ee
A few comments on this result are in order. We first realize that 
$\xi_{eff}$
entering the estimates for a successful inflation can be smaller than
the naive
$\xi$. We will discuss later whether this can improve the COBE 
constraint
(\ref{cobe}).
Note also that, if $Q_1^A=0$ (i.e. if the fields forming the inflaton 
do
not carry anomalous charge) then $\xi_{eff}=\xi$. If, on the other 
hand,
$Q_1^\alpha=0$ [i.e. if these fields do not carry charges under the
additional $U(1)$'s], then $\xi_{eff}\rightarrow 0$, and we recover the
vacuum shifting
phenomenon described in the previous section. Thus we see how the 
presence
of additional, non-anomalous $U(1)$'s can play a very significant role 
in
preventing the relaxation of the potential to zero in these 
inflationary
models.

For a viable inflationary model we should ensure that the one-loop
potential is appropriate to give a slow roll along the inflaton 
direction.
Thus,  we must consider the one-loop corrections proportional to the
Yukawa couplings introduced in the terms of eq.~(\ref{wf}).
The field-dependent masses for the scalar components of the chiral 
fields
$\Phi_\pm$ along the inflaton direction are
\bea
m^2_\pm&=\lambda^2\langle I'\rangle^2 + g_A^2 Q^A_\pm(Q_1^A\delta S^2
+\xi)+\sum_\alpha g_\alpha^2Q^\alpha_\pm Q_1^\alpha \delta =
S^2\nonumber\\
&=\lambda^2\langle I'\rangle^2 +G_{1\pm}^2\delta S^2+g_A^2Q^A_\pm\xi
\equiv \lambda^2\langle I'\rangle^2 + g_A^2a_\pm \xi,
\eea
while the fermionic partners have masses-squared equal to 
$\lambda^2\langle
I'\rangle^2$.
For large values of the field $\langle I'\rangle$, the one-loop
potential takes the form
\be
32\pi^2\delta V_1=2g_A^2(a_++a_-)\lambda^2\langle 
I'\rangle^2\xi\left(
\log\frac{\lambda^2\langle
I'\rangle^2}{Q^2}-1\right)
+g_A^4(a_+^2+a_-^2)\xi^2\log\frac{\lambda^2\langle
I'\rangle^2}{Q^2}.
\label{v1}
\ee
In this more complicated model the
scalar direction transverse to the
inflaton gains a very large mass deep in the inflaton direction. In
addition, the gauge boson corresponding to the broken $U(1)$ symmetry 
and
one  neutralino also become massive. These fields arrange
themselves in a massive vector supermultiplet, degenerate even if
$\xi\neq 0$, and their contribution to the one-loop
potential along the inflaton direction cancel exactly. The potential of 
Eq. (\ref{v1})
can be also rewritten as a RG-improved\footnote{In doing so, a careful 
treatment
of the possibility of kinetic mixing of different $U(1)$'s is required. 
The
details of our analysis are modified in the presence of such mixing but 
the
generic results are not changed.} 
tree-level potential with gauge
couplings evaluated at the scale $\lambda\langle I'\rangle$.

The term quadratic in $\lambda\langle I'\rangle$ would spoil the 
slow-roll
condition necessary for a successful inflation, but it drops out 
because
\bea
g_A^2(a_++a_-)&=&(G_{1+}^2+G_{1-}^2)\delta S^2 + g_A^2
(Q_+^A+Q_-^A)\xi\nonumber\\
&=&-G_{1I'}^2\delta S^2 - g_A^2Q^A_{I'}\xi\propto 
G_{11}^2\delta S^2 + g_A^2Q^A_1\xi=0,
\eea
where we have made use of the $U(1)$ invariance of $I'\Phi_+\Phi_-$ to 
write
the third expression which vanishes by Eq. (\ref{s2}).

The results just described for the simplest inflaton direction  
containing
more than one field are generalizable to more complicated inflatons. 
One could
have inflatons containing more than two elementary fields while still
having only a one-dimensional flat direction. Another possibility is 
that
the flat direction has  more than one free VEV (multidimensional
inflatons). It is straightforward to verify that the results obtained 
above for two mirror
fields are generic provided the inflaton does not contain some 
subdirection
capable of compensating the FI term.

As we have noticed above, the fact that the vacuum energy driving 
inflation is proportional to an effective FI $D$-term $\xi_{eff}$ may 
help in solving the problem of the mismatch between the
 value of the Fayet-Iliopoulos  $D$-term required by the COBE 
normalization and 
the value predicted by string theories. Using the notation of Eq. 
(\ref{cc}), it is easy to show that the COBE normalization imposes the 
generic constraint $V_0^{1/4}=1.6\:c^{1/4}\:\times 10^{16}$ GeV. This 
means that --for the model under consideration-- we can obtain the 
following  lower bound on the COBE normalized value of the FI $D$-term
\begin{equation}
\sqrt{\xi}_{{\rm COBE}} =
1.6\:\left(\frac{a_{+}^2+a_{-}^2}{4\pi^2}\right)^{1/4}\times 10^{16}\: 
{\rm GeV}.
\end{equation}
If we now presume that $g_\alpha\gg g_A$ and  $|Q_{\pm}^\alpha|\sim 
\beta |Q_{1}^\alpha|$ with $\beta\gg 1$, the COBE normalized value of 
the FI $D$-term becomes enhanced by a factor $\beta^{1/2}$. Whether the 
enhancement factor is large enough to reconcile the value of the  
Fayet-Iliopoulos  $D$-term required by the COBE normalization with
the value of  string theories is  very model-dependent and should be  
checked case by case. We feel encouraged, though, by the fact that the 
presence of a plethora of fields and several non-anomalous additional 
abelian symmetries in string theories may help  in solving
 the  mismatch problem.

As the next step in complexity we now examine the case in which, 
besides
the inflaton VEVs $|S_1|$ and $|S_2|$, some other field $\varphi_i$ is
forced to take a VEV (this can be triggered by $\xi$ in the anomalous
$D$-term of the potential or by $\delta S^2$ in any $D$-term). In 
general,
the new VEV can induce further VEVs too. For simplicity, we assume that
this chain of destabilizations ends with $\langle\varphi_i\rangle$.
By minimizing the $D$-term potential, all VEVs are determined to be
\be
\delta S^2 = |S_1|^2-|S_2|^2=-{\displaystyle \frac{g_A^2}{{\rm det}
G^2}}(G_{ii}^2
Q_1^A-G_{1i}^2Q_i^A)\xi\nonumber
\ee
\be
\langle\varphi_i^2\rangle = -{\displaystyle \frac{g_A^2}{{\rm det}
G^2}}(
-G_{1i}^2Q_1^A+G_{11}^2Q_i^A)\xi ,
\ee
with ${\rm det}$ $G^2=G_{11}^2G_{ii}^2-G_{1i}^4$. The tree level
potential along this direction is
\be
V_0=\frac{1}{2}g_A^2{\displaystyle \frac{\xi^2}{{\rm det}
{}G^2}}\sum_{\alpha,\beta}
g_\alpha^2g_\beta^2Q_1^\alpha
Q_i^\beta(Q_1^\alpha
Q_i^\beta-Q_1^\beta
Q_i^\alpha)
\leq \frac{1}{2}g_A^2\xi^2.
\ee
In this background, the masses of the scalar components of $\Phi_\pm$
appearing in the superpotential (\ref{wf}) are
\be
m^2_\pm=\lambda^2\langle I'\rangle^2 + g_A^2 Q^A_\pm \langle 
D_A\rangle
+\sum_\alpha g_\alpha^2Q^\alpha_\pm \langle D_\alpha \rangle =
\lambda^2\langle I'\rangle^2 + g_A^2a_\pm \xi,
\ee
and again, one finds $a_++a_-=0$.

To illustrate the above discussion, consider the following example of a
string model \cite{AF1} that satisfies the conditions required for
$D$-term inflation, at least when we restrict the analysis to a subset 
of
the fields. The $U(1)$ charges of these fields are listed in Table~I 
(we
follow the notation of ref.~\cite{jose1} with charges rescaled). For every listed field 
$S_i$, a
"mirror" field $\overline{S}_i$ exists with opposite charges. At 
trilinear order the
superpotential is
\be
\label{sup}
W=\overline{S}_{11}(S_5\overline{S}_8+S_6\overline{S}_9
+S_7\overline{S}_{10}+S_{12}S_{13})
+S_{11}(\overline{S}_5S_8+\overline{S}_6S_9+\overline{S}_7S_{10}
+\overline{S}_{12}\overline{S}_{13}).
\ee

\begin{center}
\begin{tabular}{|c|rrrrrr|}
\hline\hline
Field   &$Q_A$&$Q_3$&$Q_4$&$Q_5$&$Q_6$&$Q_7$\\
\hline\hline
$S_5$&    $-$1&    1&    0&    0& $-$2&  2   \\
$S_6$&    $-$1&    1&    0&    1&    1&  2   \\
$S_7$&    $-$1&    1&    0& $-$1&    1&  2   \\
$S_8$&    $-$1& $-$1&    0&    0& $-$2&  2   \\
$S_9$&    $-$1& $-$1&    0&    1&    1&  2   \\
$S_{10}$& $-$1& $-$1&    0& $-$1&    1&  2   \\
$S_{11}$&    0&    2&    0&    0&    0&  0   \\
$S_{12}$&    0&    1& $-$3&    0&    0&  0   \\
$S_{13}$&    0&    1&    3&    0&    0&  0   \\
\hline
\hline
\end{tabular}
\end{center}
\noindent {\footnotesize Table I: List of non-Abelian singlet fields 
with
their charges under the $U(1)$ gauge groups. The charges of these 
fields
under $U(1)_{1,2,8,9}$ are zero and not listed.}
\vspace{0.5cm}

The role of the inflaton direction can be played by $\langle S_{11}
\overline{S}_{11}\rangle$, formed by fields with zero anomalous charge. 
However for this to be viable there should be no higher order terms in the superpotential involving just the inflaton directions fields (or 
terms involving just a single non-inflaton direction field) for these 
will spoil the $F$-flatness of the inflaton direction \footnote{Superpotential terms with 
dimension strictly greater than three terms in the superpotential are suppressed 
by inverse powers of the Planck mass and so may not be so large as to 
prevent inflation. However, given the requirement that the inflaton 
must have a very large VEV $\ge {\cal O}(\sqrt{\xi})$, only very high 
dimension terms will be acceptable.} $\langle S_{11}
\overline{S}_{11}\rangle$ must be invariant under continuous gauge 
symmetries 
and so the only symmetry capable of ensuring such $F$-flatness is a 
discrete R-symmetry.  Unfortunately we do not know whether the models 
considered have such a discrete R-symmetry and thus they may allow the 
dangerous terms. Henceforth we will ignore this problem and assume the 
dangerous terms are absent.

 The
rest of the fields in the subset of Table 1 acquire large positive 
masses deep in the
inflaton direction due to the Yukawa couplings in (\ref{sup}),
guaranteeing
the stability of the inflaton direction 
$S=S_{11}=\overline{S}_{11}$.
One-loop corrections to the inflaton potential proportional to $S^2$ 
are
absent and only the $\sim\xi^2\log S^2$
dependence remains, providing the slow-roll condition. However, the end 
of
inflation poses a problem for the present example: no set of VEVs for 
the
selected fields can give zero potential. As is well known, a flat
direction ($V=0$) is always associated with an holomorphic, gauge
invariant monomial built of the chiral fields. To compensate the 
FI-term
and give $V=0$, this monomial should have negative anomalous charge.
However, in the considered subset $Q_A=Q_7/2$ and all holomorphic,
gauge invariant monomials must have then $Q_A=0$.
To circumvent this problem we enlarge the field subset by adding an 
extra
field, $S_1$  with
${\overrightarrow{Q}}(S_1)=(Q_A;Q_\alpha)=(-4;0,1,0,0,-2)$.
It is easy to see that, for example, the flat direction
$\langle 1^3,5,6,10,\overline{13}\rangle$ can cancel the FI-term and
give $V=0$. Other flat directions exist, but clearly all of them 
involve
$S_1$. However, the superpotential (\ref{sup}) does not provide a large
mass for $S_1$ when we are deep in the flat direction. Unless higher 
order
terms in (\ref{sup}) provide a positive mass for $S_1$, the FI-term
induces a destabilization of the inflaton direction and $S_1$ is forced 
to
take a VEV:
\be
\langle S_1^2\rangle=\frac{-g_A^2}{G_{11}^2}Q_1^A\xi,
\ee
where we use the definition $G_{ij}^2=g_A^2Q_i^AQ_j^A+\sum_\alpha
g_\alpha^2 Q_i^\alpha Q_j^\alpha$.
This is not a problem in itself because the rest of the fields are 
forced
to have zero VEVs and so the potential cannot relax to zero. The 
presence
of additional $U(1)$ factors prevents the vacuum shift that was
problematic for the example of section~{\bf 4}. The value of the 
potential
in the presence of a VEV for $S_1$ is
\be
V=\frac{1}{2}g_A^2\xi^2_{eff},
\ee
with
\be
\xi_{eff}^2=\frac{\sum_\alpha g_\alpha^2 
(Q_1^\alpha)^2}{G_{11}^2}\xi^2.
\ee
The masses of the rest of the fields are also affected and read:
\be
m_i^2=\lambda_i^2\langle I_i'\rangle^2 +
\frac{g_A^2}{G_{11}^2}(Q_i^AG_{11}^2-Q_1^AG_{i1})\xi,
\ee
where $\lambda_i$ are some of the Yukawa couplings in (\ref{sup}).

In general, when all the fields in the model are included, the presence 
of
the FI term will induce VEVs for the fields with negative anomalous 
charge
which are not forced to have zero VEV by $F$-term contributions. These
non-zero VEVs will in turn induce, through other $D$-terms, non-zero
VEVs for other fields, even if they have positive anomalous charge.
Finding all the VEVs requires the minimization of a complicated
 multifield potential that includes both $F$ and $D$ contributions. It 
is
intriguing, though, that the interplay of all the various fields in the 
game
 may help in reducing the FI $D$-term. As we have noted, this suggests 
a way of
 reconciling the COBE normalized value of $\sqrt{\xi}$ with the one 
suggested by
string theory.

In many cases, as in the example of section~{\bf 3}, the field VEVs
adjust themselves to give $V=0$ and no $D$-term inflation is 
possible.
In other cases however, especially in the presence of additional $U(1)$
factors, there is a limited number of fields that must necessarily
take a VEV to cancel the FI term. If the inflaton direction provides
a large $F$-term mass for them, cancelation of the FI-term is 
prevented.
Even if many other fields are forced to take VEVs, no configuration 
exists
giving $V=0$ and $D$-term inflation can take place in principle.
To determine if that is the case, one should minimize the effective
potential for large values of the inflaton field and determine all the
additional VEVs triggered by the FI-term. These VEVs, of order $\xi$
will affect the details of the potential along the inflaton direction,
both at tree level (offering the possibility of reducing the effective
value of $\xi$) and at one-loop, via their influence on the
field-dependent masses of other fields.

 {\bf 6.} Let us now discuss the post-inflationary phenomenology of 
reheating in
$D$-term inflation in superstring inspired models.
 $D$-term inflation  is  characterized by the problem of mantaining the 
reheating
 temperature $T_R$  small enough not to overproduce dangerous relics 
such as
 gravitinos \cite{gravitino}.  In fact, this  problem is common to     any
supersymmetric hybrid  model of inflation --including the ones where 
inflation is
driven by some $F$-term  \cite{shafi,linderiotto}--with the COBE 
normalized value
of the vacuum energy $V^{1/4}$   close to the GUT scale \cite{lr}. 
Indeed, when
inflation ends,  some heavy field that during inflation is   located at 
the  origin
rolls down to the true minimum and promptly  releases the vacuum 
energy. The
reheating temperature is therefore quite large, $T_R\sim V^{1/4}\sim 
10^{15}$ GeV.
On the other hand,  for unstable gravitinos
in the mass range 100 GeV to 1 TeV, one has to  require
$T_R \lsim (10^{7}-10^{9})$ GeV~\cite{gravitino}.

 We now argue that the  gravitino  bound may be naturally  satisfied  
in $D$-term
inflation inspired by superstring theories.  
As we have shown, it is  very likely --if not unavoidable-- that the 
vacuum energy
density during superstring $D$-term  inflation is 
carried by  a  combination of fields.
After slow roll, these  fields begin to oscillate about their minima of 
the
potential, and the vacuum  energy that drives inflation is converted 
into coherent
scalar field oscillations  corresponding to a condensate of 
nonrelativistic
particles. Reheating  takes place when these particles decay into light 
fields,
which through their  decays and interactions, eventually produce a 
thermal bath of
radiation. During the epoch of coherent
oscillations the  Universe is matter dominated and the energy  trapped 
in the
condensates  decreases as the cube of the scale factor. The reheating 
temperature
is determined by  the decay time of the scalar field oscillations, 
which is given
by the inverse  of the decay width $\Gamma$ of the condensates. If 
$\Gamma$ is
smaller than the  Hubble parameter $H$, the coherent oscillations phase 
is
relatively long and the reheating temperature $T_R\simeq \sqrt{\Gamma
\Mpl}$. On the  other hand, if $\Gamma\gsim H$, oscillations decay 
rapidly, and
$T_R\sim V^{1/4}$,  corresponding to 100\% conversion of the vacuum 
energy into
radiation. Since in  $D$-term inflation the decay rate of all the 
condensate
oscillations is of order of $\sqrt{\xi}$ --much larger than the Hubble 
rate
$H\sim \xi/\Mpl$ --the oscillation energy is promptly released into 
radiation with
$T_R\sim \sqrt{\xi}$.  Notice that, at this stage, there is no sequence 
of separate
reheating processes from the different condensates since their decay 
rates are
all larger than $H$.

One should not admit defeat too soon, though. It has been shown that it 
is quite natural to have a late stage of ``thermal'' inflation 
\cite{thermal} which releases a large but controlled amount of entropy 
at the electroweak scale which solves this problem. However in the 
present case there is another natural way to
avoid this  cosmological disaster which has the merit that it does not 
invoke a different sector of the theory.  Suppose that the vacuum 
manifold is
characterized by some global accidental symmetry.  This occurs, for 
instance, if in
the true vacuum with   unbroken supersymmetry the FI $D$-term is 
cancelled by two
fields $\phi_1$ and  $\phi_2$ with equal $U(1)_A$ charge $q$,
$|\phi_1|^2+|\phi_2|^2=\xi/q$. In such a case the  vacuum manifold is a 
circle in the
$(\phi_1,\phi_2)$ plane  and the accidental symmetry is an abelian 
global symmetry.
The presence of a plethora  of fields and several non-anomalous 
additional $U(1)$'s
in   string models makes the possibility of having accidental 
symmetries very
likely \cite{casas,lr1}.  This symmetry of the vacuum manifold is only 
accidental,
in the sense that it is not  respected by the Yukawa interactions in 
the
superpotential. Therefore, once supersymmetry is broken, the vacuum 
manifold is
tilted  and the Goldstone  mode $\theta$ associated to the accidental 
symmetry gets
a mass $\widetilde{m}$ --it   becomes a pseudo-Goldstone boson.   The 
value of its
mass depends upon the details of  supersymmetry breaking. In scenarios 
in which
supersymmetry is broken in some  hidden sector and is  transmitted to 
the other
fields only gravitationally, once  expects that $\widetilde{m}\sim$ 1 
TeV. However,
if supersymmetry is broken dynamically at some scale $\Lambda$ and 
transmitted to
the other fields by gauge interactions, the mass of  the 
Pseudo-Goldstone boson
may be much larger than 1 TeV if, for instance, it  communicates with 
the hidden
sector via some $U(1)$ gauge interaction, $\widetilde{m}\sim \alpha
\Lambda$ with $\alpha$ the gauge coupling constant of the messanger 
$U(1)$.

When the Hubble parameter becomes  of the order of $\widetilde{m}$ 
--much later
than the end of inflation-- coherent  oscillations of the 
Pseudo-Goldstone mode
start.  The Universe remains cold until $H$ drops below the decay width
$\Gamma_\theta\sim \widetilde{m}^3/\xi$.  At this point   the  
condensate
oscillations  decay and the decay products start thermalizing the 
Universe again,
reheating it  up to a temperature $T_R\simeq  
\widetilde{m}^{3/2}\sqrt{\Mpl/\xi}$.
This reheating  temperature is large enough to permit successful 
nucleosynthesis
and may be  even larger than the weak scale, which will allow for 
electroweak
baryogenesis. What is relevant for us, however, is that the
release of a huge amount of entropy at late epochs will  dilute any 
products of
the previous stage of reheating, including harmful gravitinos. 
Thus,  in $D$-term inflation inspired by superstring theories,
gravitinos seem to pose  no cosmological problem\footnote{We observe 
that this way
of solving the gravitino  problem via the decay of Pseudo-Goldstone 
modes at late
epochs applies to any model  of supersymmetric hybrid inflation, 
provided the
vacuum manifold after inflation posseses accidental symmetries.}.

{\bf 7.}  In conclusion, we have illustrated through different examples
the main
complications  one has to face when trying to build up a successful
$D$-term inflationary scenario out of superstring models.  The latter 
are
usually characterized by a great number of true flat directions  along
which the FI term can be cancelled. This  
makes difficult to implement the necessary conditions for $D$-term 
inflation. 
The requirement one should impose in such a case is that the theory 
possesses  the
necessary couplings between some inflaton (usually to be  identified 
with
some flat direction) and the true flat
directions, in such a way that large values of the inflaton make 
massive
at least one field along every flat direction. However,  this is not
always possible because flat directions may be protected  by chirality
from acquiring mass by coupling to
the inflaton field. This is equivalent to say that in the context of 
string
 theory, the ``vacuum shifting''  phenomenon is operative even in the 
early
Universe, thus preventing inflation from
 taking place. In those cases in which  the ``vacuum shifting'' is not 
so
efficient,  $D$-term inflation may take place with an effective FI 
$D$-term
 whose value generally depends upon the details of the minimization of 
a
complicated multifield potential. We have also shown that
 the presence of a plethora of fields and several non-anomalous 
additional abelian
 symmetries in  string theories may  solve    the problem of
the mismatch between the value of the Fayet-Iliopoulos  $D$-term 
required by the
COBE normalization and   the value predicted by string theories.  
Primordial  gravitinos may be efficiently diluted by the large amount 
of entropy released at late epochs when a pseudo-Goldstone boson 
parametrizing almost flat direction of the vacuum manifold decays into 
light states.

\vskip 0.2cm
{\bf Acknowledgements}
\vskip 0.2cm
We would like to thank P. Binetruy and B. de Carlos for useful 
discussions. This research is  supported in part by the EEC under TMR contract
ERBFMRX-CT96-0090.

\end{document}